\documentclass[aps,prb,twocolumn,superscriptaddress,showpacs]{revtex4}
\usepackage{graphicx}% Include figure files
\usepackage{dcolumn}% Align table columns on decimal point
\usepackage{bm}% bold math
\usepackage{color}%

\begin{document}

% Designing carbon nanostructures with negative Gaussian curvature

\title{Designing rigid carbon foams}

\author{Sora Park}
\affiliation{Department of Physics and
%\address{    Department of Physics and
             Research Institute for Basic Sciences,
             Kyung Hee University,
             Seoul, 130-701, Korea}

\author{Kritsada Kittimanapun}
\affiliation{Physics and Astronomy Department,
%\address{    Physics and Astronomy Department,
             Michigan State University,
             East Lansing, Michigan 48824-2320, USA}

\author{Jeung-Sun Ahn}
\affiliation{Department of Physics and
%\address{    Department of Physics and
             Research Institute for Basic Sciences,
             Kyung Hee University,
             Seoul, 130-701, Korea}

\author{Young-Kyun Kwon
\footnote    {E-mail: ykkwon@khu.ac.kr} }%
% \homepage[home page:]{http://web.khu.ac.kr/~ykkwon/}
\affiliation{Department of Physics and
%\address{    Department of Physics and
             Research Institute for Basic Sciences,
             Kyung Hee University,
             Seoul, 130-701, Korea}

\author{David Tom\'anek
% \homepage[home page:]{http://www.pa.msu.edu/~tomanek/}
\footnote   {E-mail: tomanek@pa.msu.edu;
             Permanent address:
             Physics and Astronomy Department,
             Michigan State University,
             East Lansing, Michigan 48824-2320, USA} }
\affiliation{Department of Physics and
%\address{    Department of Physics and
             Research Institute for Basic Sciences,
             Kyung Hee University,
             Seoul, 130-701, Korea}

%\ead{tomanek@pa.msu.edu}

\date{\today}
%\begin{linenumbers}

%---------------------------------------------------------------------
\begin{abstract}
We use {\em ab initio} density functional calculations to study
the stability, elastic properties and electronic structure of
$sp^2$ carbon minimal surfaces with negative Gaussian curvature,
called schwarzites. We focus on two systems with cubic unit cells
containing 152 and 200 carbon atoms, which are metallic and very
rigid. The porous schwarzite structure allows for efficient and
reversible doping by electron donors and acceptors, making it a
promising candidate for the next generation of alkali ion
batteries. We identify schwarzite structures that act as arrays of
interconnected quantum spin dots or become magnetic when doped. We
introduce two interpenetrating schwarzite structures that may find
their use as the ultimate super-capacitor.
\end{abstract}
%---------------------------------------------------------------------


% \keywords{schwarzite, density functional theory, electronic structure}


\maketitle
%---------------------------------------------------------------------

% If in two-column mode, this environment will change to single-column
% format so that long equations can be displayed. Use sparingly.
%\begin{widetext}
%\end{widetext}

% 1. Introduction
\section{Introduction}
\label{Introduction}

Nanostructured carbon has become an intensely researched topic
since the discovery of the C$_{60}$ molecule\cite{Smalley85}, a
previously unknown carbon allotrope. Most research has focussed on
structures with zero or positive Gaussian curvature such as
fullerenes, nanotubes and graphene\cite{{Dresselhaus96},{TAP111}}.
Much less emphasis has been placed on structures with negative
Gaussian curvature%
\cite{{Mackay-minimal85},{Rode-PSSb07},%
{Petersen-schwa04},{Wang-schwa03},{Merz-schwa87},%
{OKeeffe-schwa92},{Kusner-schwa95},{Lenosky-schwarzite93},%
{Benedek-schwa98},{Vanderbilt-schwa92},{Mackay-diam91},%
{Sankey-schwa92},{Lenosky-schwa92},{Crespi-schwa07},{Terrones-schwa98}}, %
related to foams, which may be equally rigid and exhibit unusual
electronic and magnetic properties. These systems, also called
schwarzites, are space-filling contiguous structures formed of
$sp^2$ bonded carbon. They are unique in their ability to
subdivide space into two disjoint, contiguous subspaces with
labyrinthine morphology and the same infinite spatial extent.

Here we focus on two schwarzite structures with cubic unit cells
containing between 152 and 200 carbon atoms. We also postulate a
structure with 400 atoms per unit cell as a representative of a
class of interpenetrating, but disconnected schwarzite lattices.
Using {\em ab initio} density functional calculations, we
determine the equilibrium structure as well as the elastic,
electronic and magnetic properties of pristine and electron or
hole doped schwarzites. Our calculations indicate the possibility
of tuning the concentration of K and Cl atoms in order to achieve
magnetic behavior. We find that the hole-doped C$_{200}$ structure
and similar systems may behave as an array of interconnected
quantum spin dots. We discuss potential energy storage
applications of schwarzites, such as next-generation electrodes
for alkali ion batteries, and calculate the capacitance of two
interpenetrating schwarzites, which may find their use as the
ultimate super-capacitor.

% 2. Computational method
\section{Computational method}
\label{Computational}

Our {\em ab initio} calculations are based on the density
functional theory (DFT) within the local density approximation
(LDA) and use the Ceperley-Alder exchange-correlation
functional\cite{CA} as parameterized by Perdew and
Zunger\cite{PZ}. Interactions between valence electrons and ions
are treated by norm-conserving pseudopotentials\cite{TM} with
separable non-local operators\cite{KB}. Atomic orbitals with
double-$\zeta$ polarization are used to expand the electron wave
functions\cite{SPOA97,Siesta} with an energy cutoff of 210~Ry for
the real-space mesh. We use 0.02~Ry as the confinement energy
shift that defines the cutoff radii of the atomic orbitals. We
sample the small Brillouin zone by 8~$k$-points in order to
represent the Bloch wave functions for the momentum-space
integration. All geometries are optimized using the conjugate
gradient method~\cite{CGmethod}, until none of the residual
Hellmann-Feynman forces acting on any atom exceeds
$1.56\times10^{-3}$~Ry/$a_{\rm{B}}$, where $a_{\rm{B}}$ is the
Bohr radius.

% 3. Results
\section{Results}
\label{Results}

\subsection{Pristine schwarzite lattices}

% entire page. There is no need to do explicit centering.
\begin{figure}[tb]
\includegraphics[width=1.0\columnwidth]{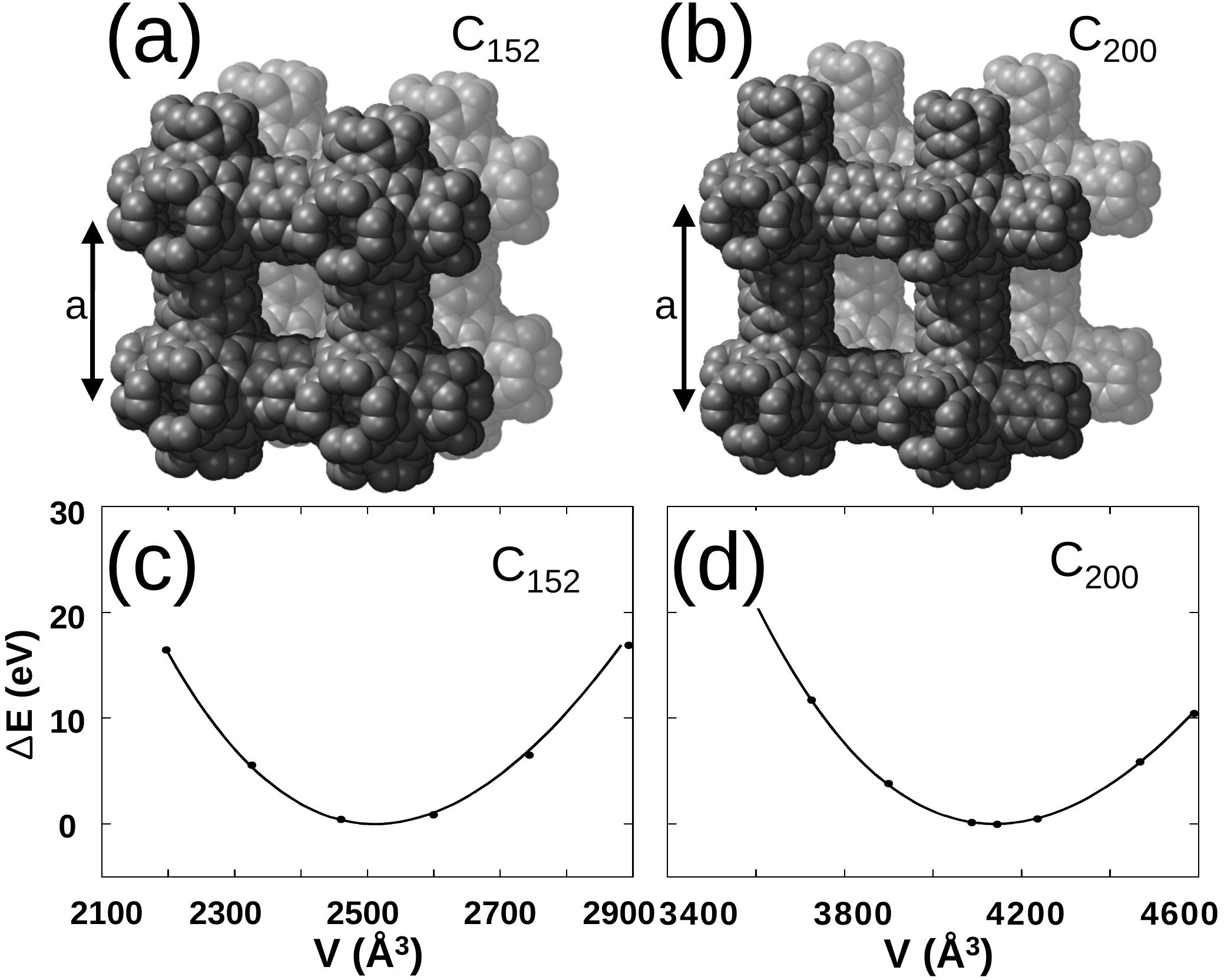}
\caption{Equilibrium geometry of $P$ schwarzites with 152 (a) and
200 (b) carbon atoms per cubic unit cell with the lattice constant
$a$ and the volume $V=a^3$. Total energy change ${\Delta}E$ as a
function of $V$ for the C$_{152}$ (c) and C$_{200}$ (d)
schwarzites. \label{Fig1}}
\end{figure}
%---------------------------------------------------------------------

Previous theoretical studies addressed schwarzites with %
24 (Ref.~\onlinecite{OKeeffe-schwa92}), %
32 (Ref.~\onlinecite{Kusner-schwa95}), %
48 (Ref.~\onlinecite{Lenosky-schwarzite93}), %
56 (Refs.~\onlinecite{Lenosky-schwarzite93},\onlinecite{Benedek-schwa98}), %
72 (Ref.~\onlinecite{Benedek-schwa98}), %
80 (Ref.~\onlinecite{Benedek-schwa98}), %
168 (Refs.~\onlinecite{Vanderbilt-schwa92}, \onlinecite{Sankey-schwa92}), %
192 (Refs.~\onlinecite{Mackay-diam91}, \onlinecite{Sankey-schwa92}), %
200 (Ref.~\onlinecite{Lenosky-schwarzite93}), %
216 (Refs.~\onlinecite{Sankey-schwa92}, \onlinecite{Lenosky-schwa92}), and %
224 (Ref.~\onlinecite{Crespi-schwa07})
atoms per unit cell, or related general morphology issues %
(Refs.~\onlinecite{Crespi-schwa07}, \onlinecite{Terrones-schwa98}). %
We focus on two structures with a primitive ($P$) minimal surface
spanned by an underlying simple cubic lattice, which have not been
studied previously and which contain 152 and 200 carbon atoms per
unit cell. The C$_{152}$ structure is depicted in
Fig.~\ref{Fig1}(a) and the C$_{200}$ structure in
Fig.~\ref{Fig1}(b). The unit cells of either schwarzite contain
one ``core'' structure with $O_h$ symmetry and a negative local
Gaussian curvature. The six extremities of the cores are connected
to neighboring cores by $(4,4)$ carbon nanotube segments of
various length to form a contiguous lattice. Negative Gaussian
curvature is introduced by the presence of heptagons in the
graphitic lattice. Our systems contain exactly 24 heptagons per
unit cell in accordance with Euler's theorem that, along with 24
hexagons, form the cores of the schwarzite. Each interconnect
contains one nanotube unit cell with 8 hexagons in the C$_{152}$
structure and two unit cells with 16 hexagons in the C$_{200}$
structure.

To identify the equilibrium structure and elastic properties of
the two schwarzite structures, we performed a series of structure
optimizations at fixed unit cell volume. The total energy change
${\Delta}E(V)$ with respect to the optimum structure is presented
for the two structures in Figs.~\ref{Fig1}(c) and \ref{Fig1}(d)
along with Murnaghan function fits of our data. For the C$_{152}$
schwarzite we find the binding energy $E_{coh}=10.50$~eV/atom at
the equilibrium volume $V_{eq}=2515$~\AA$^3$, corresponding to the
lattice constant $a=13.60$~{\AA}, gravimetric density
${\rho}=1.21$~g/cm$^3$ and bulk modulus $B=111$~GPa. The less
dense C$_{200}$ schwarzite structure is slightly more stable with
$E_{coh}=10.52$~eV/atom at $V_{eq}=4141$~\AA$^3$, corresponding to
$a=16.06$~{\AA}, but has a smaller gravimetric density
${\rho}=0.96$~g/cm$^3$ and bulk modulus $B=76.4$~GPa. The
reduction of the binding energy with respect to the value of
10.88~eV/atom in a graphene monolayer is caused by the strain
associated with curvature and is similar in value to carbon
nanotubes. We find both structures to be substantially lighter
than graphite (${\rho}=2.27$~g/cm$^3$) and diamond\cite{Kittel05}
(${\rho}=3.54$~g/cm$^3$). Also the bulk modulus of the stiffer
C$_{152}$ is about four times smaller than the diamond
value\cite{Kittel05} $B_{expt}=442$~GPa.

% entire page. There is no need to do explicit centering.
\begin{figure}[tb]
\includegraphics[width=1.0\columnwidth]{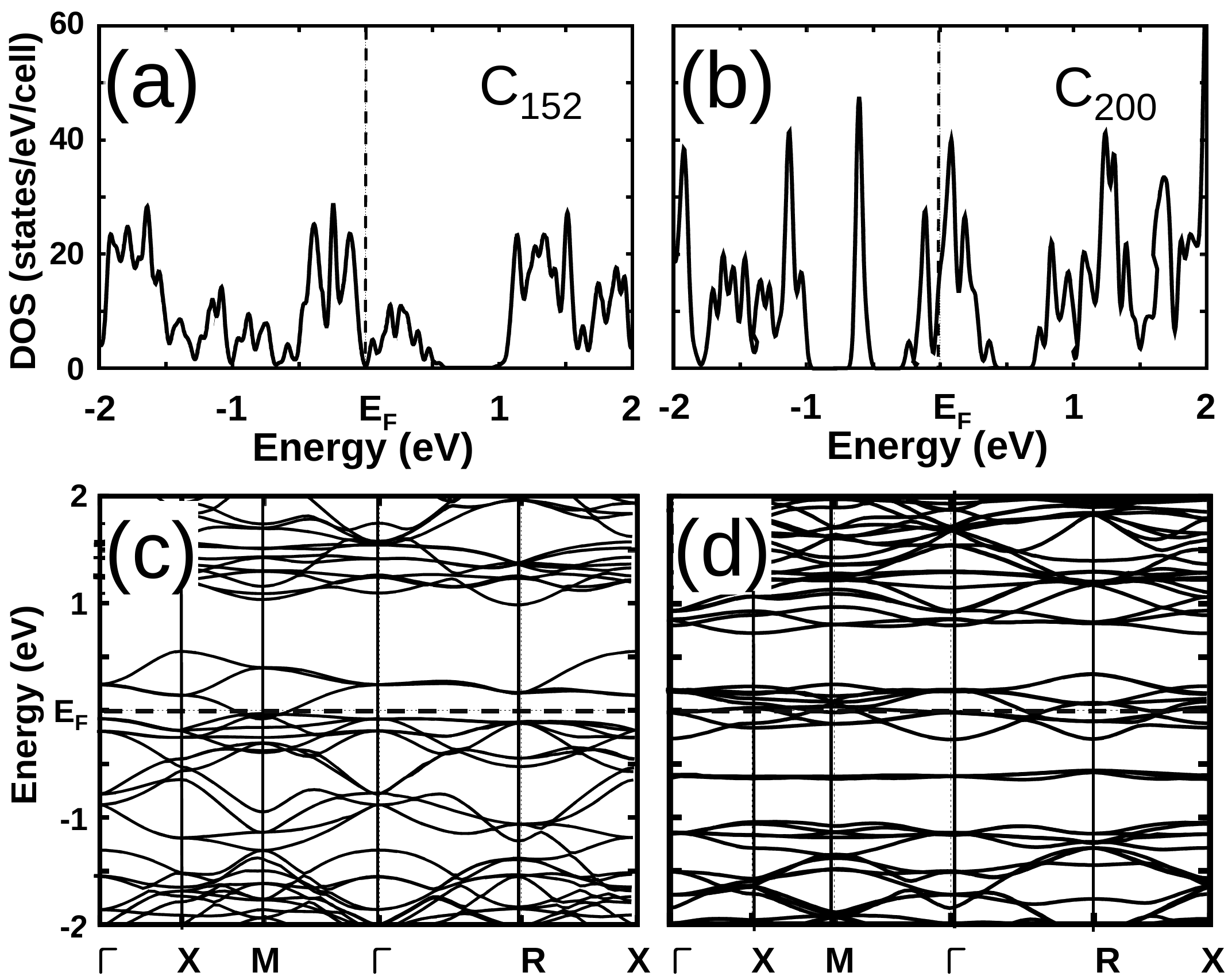}
\caption{Electronic structure of schwarzites. Density of states of
the optimized schwarzite structures with unit cells containing 152
(a) and 200 (b) C atoms, smoothened by a Gaussian with the full
width at half maximum of 0.03~eV. Corresponding band structure of
the C$_{152}$ (c) and C$_{200}$ (d) schwarzites. \label{Fig2}}
\end{figure}
%---------------------------------------------------------------------

To characterize the electronic properties of these schwarzites, we
calculated their electronic density of states and band structure.
The electronic densities of states of C$_{152}$ and C$_{200}$,
presented in Figs.~\ref{Fig2}(a) and \ref{Fig2}(b), indicate that
both structures are metallic. The origin of the individual peaks
in the density of states can be traced back to the band structure
graphs in Figs.~\ref{Fig2}(c) and \ref{Fig2}(d). The schwarzites
exhibit narrow band gaps above and, in the case of C$_{200}$, also
below the Fermi level. Furthermore, as seen in Figs.~\ref{Fig2}(b)
and \ref{Fig2}(d), the C$_{200}$ structure exhibits a 3-fold
degenerate flat band about 0.62~eV below the Fermi level. Since in
neither structure the Fermi level falls into a sharp peak, both
structures are non-magnetic.

% entire page. There is no need to do explicit centering.
\begin{figure}[t]
\includegraphics[width=1.0\columnwidth]{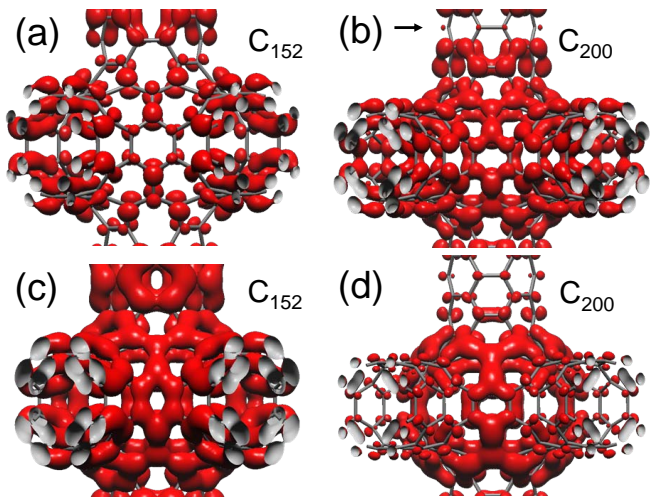}
\caption{(Color online) Electron densities in schwarzite
structures containing 152 and 200 atoms per unit cell, presented
as isodensity surfaces. Electron density associated with frontier
states of C$_{152}$ (a) and C$_{200}$ (b) in the energy range
$E_F-0.15$~eV$<E<E_F+0.15$~eV, presented at the isodensity value
of $10^{-2}$~{\AA}$^{-3}$. (c) Electron density of C$_{152}$
associated with states in the energy range
$E_F-0.72$~eV$<E<E_F+0.10$~eV, presented at the isodensity value
of $10^{-2}$~{\AA}$^{-3}$. (d) Electron density of C$_{200}$
associated with the localized state near $E_F-0.62$~eV, presented
at the isodensity value of $10^{-4}$~{\AA}$^{-3}$.
% EF(C152) = -4.38 eV; EF(C200) = -3.86 eV
\label{Fig3}}
\end{figure}
%---------------------------------------------------------------------

Besides the nonzero value of the density of states at $E_F$ in the
schwarzites we study, which makes them metallic, also the nature
of those states is very important for the dielectric response of
the system. Therefore, we plot the electron density associated
with the frontier states at the Fermi level of C$_{152}$ in
Fig.~\ref{Fig3}(a) and that of C$_{200}$ in Fig.~\ref{Fig3}(b). We
find the frontier states to be highly delocalized in both systems.
In the larger C$_{200}$ schwarzite, we note that the frontier
states have a node in the middle of the tubular interconnects
between the cores, indicated by an arrow in Fig.~\ref{Fig3}(b). As
seen in Fig.~\ref{Fig3}(c), which depicts the electron density of
C$_{152}$ associated with the ${\approx}0.8$~eV wide band mostly
below the Fermi level, the corresponding states are distributed
nearly uniformly across the structure. To understand the origin of
the flat band in the C$_{200}$ schwarzite 0.62~eV below the Fermi
level, we present the associated electron density in
Fig.~\ref{Fig3}(d). We may characterize this state as an array of
non-interacting quantum dots centered in the negative Gaussian
curvature regions at the cores of the C$_{200}$ structure. We
expect such a state to occur also in other schwarzite structures
of the same type, which have larger unit cells due to longer
tubular interconnects.

\subsection{Interpenetrating schwarzite lattices}

% entire page. There is no need to do explicit centering.
\begin{figure}[t]
\includegraphics[width=1.0\columnwidth]{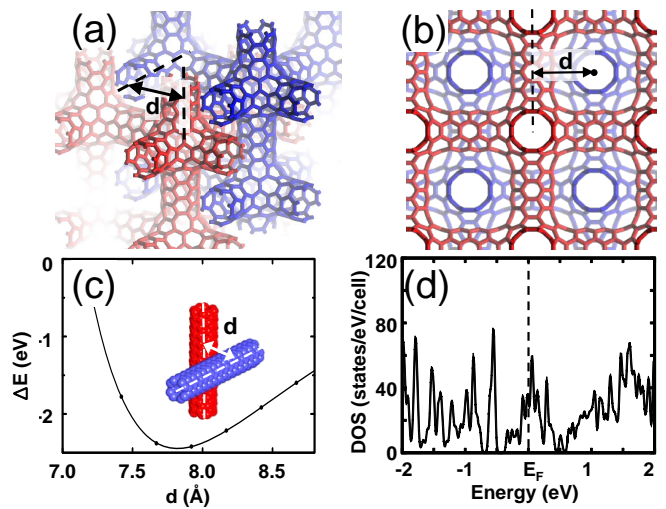}
\caption{(Color online) Equilibrium structure of two
interpenetrating C$_{200}$ schwarzite lattices, shown in
Fig.~\protect\ref{Fig1}(b), with 400 C atoms in total per unit
cell, shown in perspective (a) and top (b) view. (c) Interaction
energy between two $(4,4)$ nanotubes, with the geometry shown in
the inset, representing the interaction between the tubular
interconnects in the two interpenetrating C$_{200}$ schwarzites.
(d) Electronic density of states of the C$_{400}$ schwarzite.
\label{Fig4}}
\end{figure}
%---------------------------------------------------------------------

The open space in the C$_{152}$ and C$_{200}$ structures, depicted
in Figs.~\ref{Fig1}(a) and \ref{Fig1}(b), can be filled by other
atoms as external dopants. As we show in the following and depict
in Figs.~\ref{Fig4}(a) and \ref{Fig4}(b), the open space in the
C$_{200}$ schwarzite is wide enough and ideally suited to
accommodate even an identical replica of the C$_{200}$ schwarzite
in an unstrained lattice.

In either schwarzite described in this manuscript, the tubular
interconnects are composed of $(4,4)$ carbon nanotubes. As seen in
Figs.~\ref{Fig4}(a) and \ref{Fig4}(b), the closest approach
distance between the sublattices occurs in the geometry depicted
in the inset of Fig.~\ref{Fig4}(c), with two $(4,4)$ nanotube
segments standing normal to each other. To estimate the optimum
inter-wall separation, we calculated the inter-tube interaction in
this geometry and plotted the data points along with a Morse
function fit in Fig.~\ref{Fig4}(c). We found the optimum
inter-tube distance to be $d_{eq}=7.8$~{\AA}, corresponding to the
closest inter-wall distance of $2.4$~{\AA}, somewhat smaller than
the inter-layer distance in graphite. The corresponding distance
between adjacent tubular interconnects of the C$_{200}$
substructures in C$_{400}$, as depicted in Figs.~\ref{Fig4}(a) and
\ref{Fig4}(b), is $d=8.00$~{\AA}, which indeed is very close to
the optimum inter-tube distance value $d_{eq}=7.8$~{\AA}. Since
the C$_{400}$ schwarzite is structurally a superposition of two
C$_{200}$ schwarzite lattices, its gravimetric density is twice
that of C$_{200}$, namely ${\rho}=1.92$~g/cm$^3$. Also the bulk
modulus of C$_{400}$, $B=153$~GPa, is twice that C$_{200}$, which
is about one third of the diamond value.

As the electronic density of states of the C$_{200}$ sublattices,
also that of the C$_{400}$ schwarzite, shown in
Fig.~\ref{Fig4}(d), has a large density of states at the Fermi
level, indicating a metallic system. Similar to C$_{200}$, also
C$_{400}$ has a narrow band about $0.6$~eV below $E_F$, associated
with a lattice of very weakly coupled quantum dots. Careful
comparison of Figs.~\ref{Fig2}(b) and \ref{Fig4}(d) indicates that
the central band around $E_F$ becomes wider in the C$_{400}$
lattice in comparison to the C$_{200}$ lattice. This band
broadening is accompanied by a corresponding narrowing of the band
gaps above and below this central band.

\subsection{Intercalated schwarzites}

As mentioned above, the empty space delimited by the schwarzite
minimal surface can accommodate intercalant ions in order to shift
the Fermi level. We consider K as a model donor that has been used
widely in graphite intercalations compounds
(GICs)\cite{Dresselhaus-GIC02}. Due to its highest electronic
affinity in the periodic table\cite{Kittel05} of $3.61$~eV, we
choose Cl as a model acceptor.

As in the previous subsection, we focus on the C$_{200}$
schwarzite structure. To maximize charge transfer between the
intercalant atoms and the schwarzite lattice, we place one K or Cl
atom in the center of each core of the schwarzite. To estimate the
charge transfer between the intercalants and the schwarzite, we
performed a Mulliken population analysis and, alternatively,
integrated the total electron density within a sphere surrounding
the intercalant ion. We found the Mulliken charge on K to be
$+0.96e$ and that on Cl to be $-0.42e$. Electron density
integration around the intercalant within schwarzite only makes
sense up to the size of the cavity inside the core structure.
Considering a sphere with a radius of $R=2.4$~{\AA} around Cl
yields $7.00$ valence electrons in case of an isolated atom and
$7.42$ valence electrons in case of Cl@C$_{200}$. Since the atomic
radius of K is much larger, an isolated atom contains only $0.56$
valence electrons within a sphere with $R=2.4$~{\AA}. Upon
intercalation, in K@C$_{200}$, it loses almost all of the valence
charge, so that only $0.01$ valence electrons remain within this
sphere.

By subtracting the total energy of C$_{200}$ and the isolated atom
A from that of A@C$_{200}$, where A$=$K or Cl, we found that the K
atom stabilizes the unit cell by $2.080$~eV, whereas the Cl atom
provides a smaller stabilization by $0.636$~eV per unit cell,
partly due to the smaller charge transfer.

% Total energies SIESTA (DT notebook 11 Nov 09, p 117)
% Isolated atoms:
% Etot(C)=-144.373 eV
% Etot(K)=-7.897 eV
% Etot(Cl)=-403.740 eV
% Pristine structures:
% Etot(C200)=-30978.262 eV
% Etot(C400)=-61971.988 eV
% => Etot(C400)-2*Etot(C200)=-15.464 eV
% Doped structures:
% Etot(K@C200)=-30988.240 eV
% Etot(Cl@C200)=-31382.638 eV
% Etot(K+Cl@C400)=-62386.613 eV
%
% Etot(Cl@C200)-Etot(C200)-Etot(Cl)=-0.636 eV
% Etot(K@C200)-Etot(C200)-Etot(K)=-2.080 eV
% Etot(K+Cl@C400)-2*Etot(C200)-Etot(K)-Etot(Cl)=-2.987 eV
% Etot(K+Cl@C400)-Etot(C400)+2*Etot(C200)-Etot(K@C200)-Etot(Cl@C200)=-0.271 eV
%
% Charge integration inside a sphere of radius R
%
% Isolated Cl, R=4.5 a.u.: Q=6.999e
% Cl@C200, R=4.5 a.u.: Q=7.417e
%
% Isolated K, R=4.5 a.u.: Q=0.559e
% K@C200, R=4.5 a.u.: Q=0.005e
%
% Isolated Cl, R=4.73 a.u. = 2.5 A: Q=7.000e
% Cl@C200, R=4.73 a.u. = 2.5 A: Q=7.420e
%
% Isolated K, R=4.73 a.u. = 2.5 A: Q=0.618e
% K@C200, R=4.73 a.u. = 2.5 A: Q=5.885e
%
% Isolated Cl, R=5.67 a.u. = 3.0 A: Q=7.000e
% Cl@C200, R=5.67 a.u. = 3.0 A: Q=7.502e
%
% Isolated K, R=5.67 a.u. = 3.0 A: Q= 0.812e
% K@C200, R=5.67 a.u. = 3.0 A: Q=0.193e

% Electron doping

The ability to efficiently transfer most of the valence charge
from the alkali atom to the graphitic structure is used currently
in Li ion batteries, where Li is intercalated in-between graphene
layers in graphite, causing an ${\approx}30$\% expansion in the
direction normal to the graphene layers. Repeating expansion and
contraction during charge/discharge cycles is known to reduce the
lifetime of alkali ion batteries, as it causes structural
rearrangement of the grains and eventually blocks diffusion
pathways. No such adverse effects should occur in the rigid, but
still conductive schwarzite structures, which may yield a new
generation of advanced alkali ion batteries. As a different
potential application of doped schwarzites, we wish to mention
their potential for hydrogen storage, similar to Ca doped C$_{60}$
crystals that proved to store reversibly molecular
hydrogen\cite{MNYoonC60GIC08} up to 8.4\%(wt).

% Hole doping

As suggested by the smaller amount of charge transferred from the
C$_{200}$ schwarzite to Cl ions, the effect of acceptor
intercalation is much smaller than that of donor intercalation.
Nevertheless, hole doping should make it possible to shift $E_F$
down. Especially interesting appears the possibility to move the
Fermi level into the region of the threefold degenerate flat band.
Partial occupation of the band is expected to change the system to
a ferromagnet with a magnetic moment per unit cell up to
$6{\mu}_B$. To determine the necessary amount of doping, we
estimated the occupied portion of the band near $E_F$ to carry $6$
electrons. Combining this with the fact that the flat band also
contains $6$ electrons, we conclude that hole doping C$_{200}$
with $7-11$ electrons per unit cell should, assuming that the
rigid band model applies, move the Fermi level into the region of
the flat band. Using the charge transfer estimated for one Cl atom
in C$_{200}$, this could be achieved by intercalating between
${\approx}10-30$ Cl atoms per unit cell. Using interatomic
distances from free Cl$_2$, we believe that in the optimum case,
Cl$_6$ clusters could be accommodated in the cores. Assuming that
also each tubular interconnect could accommodate up to two Cl
atoms, the Cl doped system should barely reach the criterion of
partly depleting the initially flat band by Cl doping, causing
ferromagnetism to occur.

% entire page. There is no need to do explicit centering.
\begin{figure}[t]
\includegraphics[width=1.0\columnwidth]{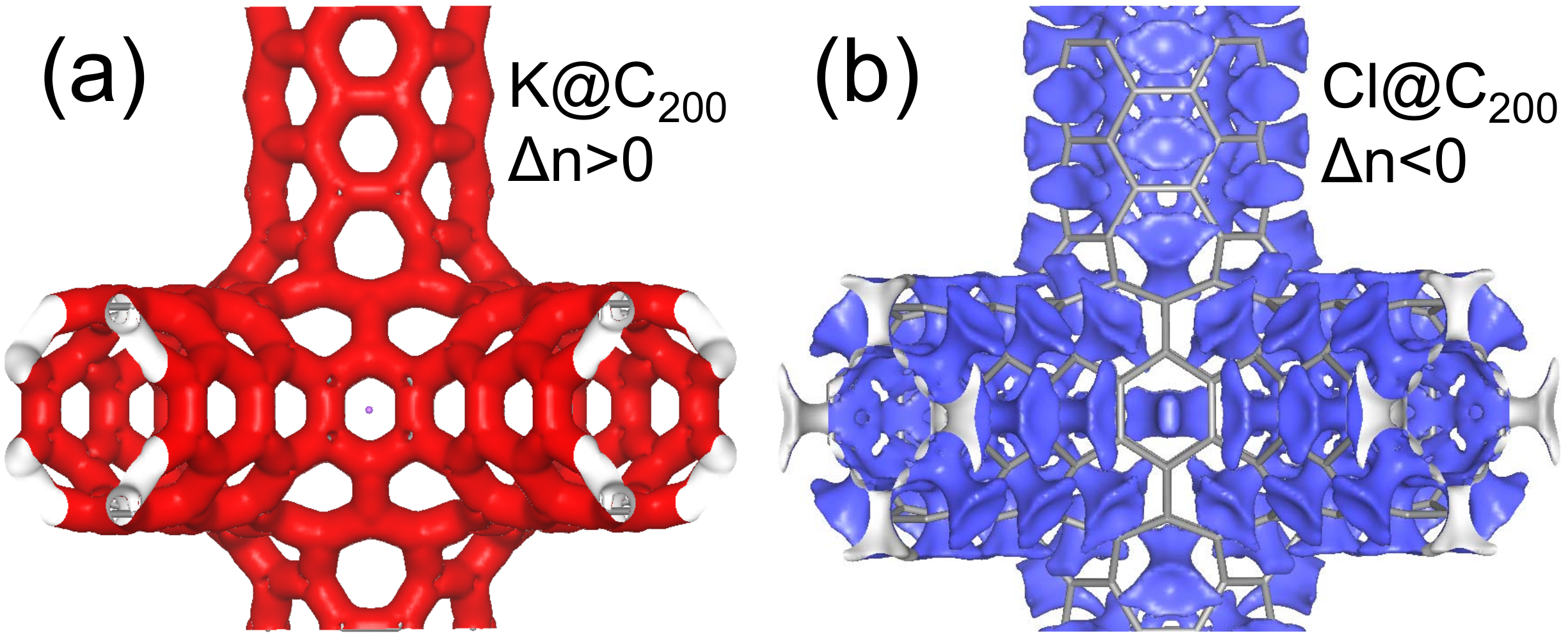}
\caption{(Color online) Electron density difference
${\Delta}n(A@C_{200}) =
n_{tot}(A@C_{200})-n_{tot}(A)-n_{tot}(C_{200})$, providing insight
into the charge flow in the C$_{200}$ schwarzite following
intercalation with $A=$ K, Cl atoms. (a) ${\Delta}n(K@C_{200})$ at
the isodensity value $+0.04~${\AA}$^{-3}$. (b)
${\Delta}n(Cl@C_{200})$ at the isodensity value
$-0.04$~{\AA}$^{-3}$. \label{Fig5}}
\end{figure}
%---------------------------------------------------------------------

% Electron density difference

To get an impression of how uniformly the transferred charge is
distributed across the schwarzite lattice, we plot in
Fig.~\ref{Fig5}(a) the distribution of the excess electron density
in K@C$_{200}$ and in Fig.~\ref{Fig5}(b) the corresponding
distribution of the deficient electron density in Cl@C$_{200}$.
Since both ions are well separated from the schwarzite lattice, we
find that the transferred charge is distributed rather uniformly
across the schwarzite in both cases.

% C400 as two interpenetrating lattices

Also the two interpenetrating C$_{200}$ lattices, forming the
C$_{400}$ schwarzites, can be differentially doped, one with K and
the other with Cl atoms, giving a net positive charge to one and a
net negative charge to the other sublattice, similar to a
capacitor. We superposed the K@C$_{200}$ and Cl@C$_{200}$ lattices
and found a net stabilization with respect to separated
K@C$_{200}$ and Cl@C$_{200}$ doped schwarzites of $-15.735$~eV.
Similar to the doped sublattices, the Mulliken population analysis
indicates that the net charge on K is close to $+0.70e$ and that
on Cl is close to $-0.42e$. The opposite charges are then
accommodated in the adjacent C$_{200}$ substructure.

In the following we will estimate the capacitance of the
schwarzite structure per $1$~cm$^3$ volume. The capacitance of a
parallel-plate capacitor of area $A$ and inter-plate distance $d$,
defining the volume $V=Ad$, is given by $C={\epsilon_0}A/d$ in SI
units. For the sake of reference, an idealized capacitor with
$V=1$~cm$^3$, formed by a graphene bilayer with $d=0.335$~nm,
would have the area $A=3.0{\times}10^3$~m$^2$ and a capacitance of
$79.3$~F. In the unit cell of the C$_{400}$ schwarzite with the
volume $V=4.141$~nm$^3$, 200 carbon atoms are associated with one
electrode. With an estimated area per carbon atom of
$2.62$~{\AA}$^2$, same as in graphene, we estimate that the area
of one electrode per cm$^3$ schwarzite should be
$1.3{\times}10^3$~m$^2$. Assuming furthermore the same average
interlayer distance as in a graphene bilayer, we estimate the
capacitance of the C$_{400}$ schwarzite to be $34.3$~F per cm$^3$
material.

As an independent way to estimate the capacitance of the C$_{400}$
schwarzite, we carefully inspected the total energy differences
between neutral C$_{200}$ and C$_{400}$ systems, as well as the
total energies of K@C$_{200}$, Cl@C$_{200}$ and C$_{400}$
containing both K and Cl. The latter structure, which contains one
positively and one negatively charged C$_{200}$ minimal surface
per unit cell, is a superposition of K@C$_{200}$ and Cl@C$_{200}$.
The stabilizing interaction between the charged C$_{200}$
substructures of $15.735$~eV is partly due to the chemical
interaction between the two electrodes, which we estimated to be
$15.464$~eV per unit cell in the pristine system. We assign the
remaining part of the interaction energy, $0.271$~eV, to be the
energy associated with the electric field between the capacitor
plates. To understand the meaning of this energy gain, we consider
two parallel plate electrodes carrying constant charges $+Q$ and
$-Q$, approaching from infinity to form a parallel-plate capacitor
of capacitance $C$, thereby gaining the energy
${\Delta}U=Q^2/(2C)$. Considering ${\Delta}U=0.271$~eV and
$|Q|=0.42e$, the smaller of the charges transferred between the
intercalant atoms and the neighboring electrodes, we can estimate
the capacitance $C$ per unit cell and the corresponding value in
the bulk C$_{400}$ schwarzite material, which turns out to be
$12.6$~F per cm$^3$.

We would like to point out that the above capacitance values are
based on rough approximations and thus should be considered as
order-of-magnitude estimates. Still, we are pleased that our
estimate based on total energy differences in the intercalated
system lies close to the estimate based on area and inter-plate
distance. In any case, the estimated capacitance exceeds that of
nowadays' capacitors by several orders of magnitude. Should a
material close to the postulated C$_{400}$ schwarzite ever be
synthesized, we must also consider the need to connect the two
interpenetrating sublattices to two leads, which will be a
nontrivial task. We also need to point out that the close
proximity of the two electrodes in the schwarzite material
promotes tunneling, which would eventually discharge the
electrodes over time. Still, the expected capacitance of tens of
Farads per cubic centimeter is an appealing prospect.

In summary, we studied the stability, elastic properties and
electronic structure of $sp^2$ carbon periodic minimal surfaces
with negative Gaussian curvature, called schwarzites, using {\em
ab initio} density functional calculations. Our studies of two
primitive minimal surfaces, spanned by an underlying simple cubic
lattice with 152 and 200 carbon atoms, indicate that these systems
are very stable, rigid, and electrically conductive. The porous
schwarzite structure allows for efficient and reversible doping by
intercalation of electron donors or acceptors. We identified
systems that should act as arrays of interconnected quantum spin
dots or, when doped, exhibit ferromagnetic behavior. We also
introduce two interpenetrating schwarzite structures as an unusual
system that may find its use as the ultimate super-capacitor.

%Acknowledgements

We
% thank Jeung-Sun Ahn for useful discussions and
acknowledge contributions of Thomas Moore and Daniel Enderich to
the visualization of schwarzite structures. DT was supported by
the International Scholar fellowship at Kyung Hee University and
funded by the National Science Foundation under NSF-NSEC grant
425826 and NSF-NIRT grant ECS-0506309. YK was supported by NRF of
Korea grant KRF-2009-0074951.

\quad \\

%+++++++++++++++++++++++++++++++++++++++++++++++++++++++++++++++++++++
% \bibliographystyle{unsrt}% your bst file here
% \bibliographystyle{apsrev}% your bst file here
% \bibliography{schwa09-notitles} %your bib file here
% \end{document}
%+++++++++++++++++++++++++++++++++++++++++++++++++++++++++++++++++++++

\end{document}